# A Physicist's view on Chopin's Études

Massimo Blasone

Dipartimento di Fisica & INFN, Università di Salerno,

Via Giovanni Paolo II, 132, 84084 Fisciano (SA), Italy

Email: mblasone@unisa.it

## Abstract

We propose the use of specific dynamical processes and more in general of ideas from Physics to model the evolution in time of musical structures. We apply this approach to two Études by F. Chopin, namely op.10 n.3 and op.25 n.1, proposing some original description based on concepts of symmetry breaking/restoration and quantum coherence, which could be useful for interpretation. In this analysis, we take advantage of colored musical scores, obtained by implementing Scriabin's color code for sounds to musical notation.

## 1 – Introduction

The relationship between Music and Mathematics has a very long history [1,2], starting perhaps with Pythagoras who first associated musical intervals with rational numbers, till our days where it represents an active field of research for mathematicians, computer scientists, physicists and musicians.

However, apart from the application of mathematical models and algorithms in the production of music (an example is fractal music, see Ref.[3]), such studies seem to have little impact on the musical praxis. In particular, for what concerns Classical Music, it exists a consolidated tradition for musical performance which apparently takes no advantage in the knowledge of the mathematical structure underlying music. This situation is somehow similar to that of a (al)chemist who knows the rules to combine elements for obtaining a given substance and does not really care of the complex (quantum) dynamics which rules atoms and molecules.

Of course, there is also another, deeper reason for this attitude: we do music for our pleasure, it represents an efficient and universal way of communicating our emotions. Thus it is certainly not enough to study the mathematical structures hidden in musical language without also considering the level of perception, which is where these structures are eventually elaborated (and appreciated). In other words, a beautiful mathematical form is not necessarily associated to beautiful music (in the artistic sense) and vice-versa.

Thus one could be tempted to dismiss these studies as interesting but not really able to touch the level of *aesthetic experience*, which is the one we are actually interested in when performing/listening music.

Nevertheless, it is surely a common experience to everybody playing a page by Bach or Mozart, the deep sense of wonder in front of these masterpieces and the feeling that should be "something there" associated to what we perceive as Beauty. This is even stronger, we think, when the musician is also a scientist because then he/she will inevitably try to search for the laws underlying such beautiful structures.

This is indeed the main motivation for the considerations contained in this work: they are not the result of the application of some theoretical program to a piece of music; rather they arise when sitting at the piano trying to understand the hidden logic of some compositions with the aim of properly performing them, which means efficiently transmitting their information content to ourselves and to other people.

A typical way of talking about Music (and more in general about emotions) is by use of metaphors, images or situations, which help to better understand the musical language [4]. This is particularly evident in the case of programmatic music which by definition refers to some specific subject. In any case, such associations are necessarily vague since they are linked to the music via the effects they both (music and image) produce in our brains, which of course are highly subjective.

The idea is therefore to use as "physical metaphors" of a given music some specific dynamical processes, which can then be in turn associated to a variety of physical processes (images). The advantage is that such dynamics can be directly related to the mathematical/tonal structure of the music under consideration: one can thus apply concepts as symmetry, forces, (dis)order, phase transitions, coherence, etc. which have a well-defined meaning in Physics.

Of course, in doing this it is necessary to proceed with much care and with a certain degree of approximation, maintaining a *qualitative approach* in order to avoid the danger of being "reductionist", i.e. to pretend that Music could be entirely described in mathematical terms. Another risk, which is in some sense unavoidable, is that of "manipulating" the music, by forcing some interpretation which is not entirely appropriate.

This approach can be in principle applied to any music. Here we consider Chopin's Études for various reasons: one is very personal and has to do with the preference for this Author and in particular for his Études and Preludes. There are however also some more objective motivations: the simple form of these compositions (miniatures, in the case of some preludes) much less structured with respect to other forms (like the Sonata for example), makes relatively easy their analysis; also, their character is very evocative of some (natural) images, as pointed out by many authors/editors (see Section 2). Another aspect of Chopin's music is its universality, being appreciated by a very large number of people, independently of their cultural background. Finally, a certain enigmatic character of this Author, as was recognized already from his contemporaries (Le Pianiste, 1833), together with the fact that he refused any programmatic interpretation for his music, makes particularly intriguing its analysis.

The paper is organized as follows: in Section 2 we give some historical notes on Chopin's Études; Sections 3 and 4 are devoted to the analysis of the Études Op.10 n.3 and Op.25 n.1, respectively; Section 5 contains conclusions and perspectives of this work. In a separate appendix, we illustrate the implementation of the color code by Scriabin for musical tones, which is used for the analysis of Sections 3 & 4.

## 2 – Some historical notes on Chopin's Études

The term étude historically denotes an instrumental composition written for didactic purposes, for the aim of practicing some particular combination of notes or a specific movement. Such compositions have typically a simple structure and are rather short with respect to other more structured forms like for example the Sonata. Early examples of études for keyboard are the *Essercizi per gravicembalo* ("30 Exercises for harpsichord", 1738) by D.Scarlatti, and later the collections by J.B.Cramer, M.Clementi, C.Czerny and I.Moscheles to name just a few.

Fryderyk Chopin composed two sets of études: 12 Études Op.10, published in 1833 and 12 Études Op.25, published in 1837; he later published 3 Études for the Moscheles method and some of the 24 Preludes (published in 1839) are indeed very similar to the Études op. 10 and 25.

Chopin's Études are devoted to the development of specific technical aspects (e.g. Op.10 n.2 for chromatic scales, Op.25 n.10 for octaves, Op.25 n.6 for double thirds, etc.) pushed to the highest peaks of difficulty (at least for that epoch). This intent is confirmed by the dedication of Op.10 to the great piano virtuoso F.Liszt (Op.25 is dedicated to M. d'Agoult, who was the mistress of Liszt). It is also documented [5] that Chopin wrote some of the études Op.10 as exercises for himself in practicing his Concerto n.1 for piano and orchestra. Another input was represented by the 10 concerts given by N.Paganini in Warsaw in 1829, which the young Chopin had occasion to listen. As many of his contemporaries (including Liszt), he was deeply impressed by the technique and personality of the Italian violinist: Paganini's 24 Capriccios for the violin were a model for Chopin's Études and Preludes (together with the Well-Tempered Clavier by J.S.Bach).

In Chopin, the technical aspect is however never limited to didactic purposes, rather it becomes instrumental for expressing a large variety of emotions: he was able to elevate the études from the status of exercises to that of piece of Art, creating a new genre – the concert étude – which was then further developed by Liszt, Debussy, Rachmaninov and others. This achievement is well synthesized in a sentence by L.Kentner, for which Chopin's Études represent a "perfect fusion of the athletic and aesthetic" [6]: there, technique is developed in its etymological sense of τέχνη, a word that for ancient Greeks had also the meaning of "Art".

These compositions, which still today represent a summa of piano virtuosity, presented novelties which were surely revolutionary for that time: very large positions for the hand, new ways of attacking the keyboard, together with novel harmonic combinations, large harmonic progressions, instrumental figurations used as themes (cantabile character of figuration). In some respect, the innovative harmonic treatments in Chopin's Etudes anticipate the dissolution of tonality [7].

For their very original character, these features were often not understood by his contemporaries and attracted to him some criticism. Perhaps the most known is the one by Berlin editor and composer L.Rellstab who wrote about the Op. 10: *"Those who have crooked fingers, can treat them by means of these exercises. But those who suffer from no such ailment would do well to avoid them"* [6]. Also F.Mirecki criticized the monotony (!) of the themes in the Études, saying that *"if notes would be colored, instead of black, these scores could be used as wallpaper"* [6].

On the other hand, others immediately recognized the great value of this music: the Études op.10 were very positively reviewed in the magazine *Le Pianiste*, just after their publication in 1833. Also R.Schumann, as a music critic, praised highly the young composer (*"Hats off, gentlemen, a genius!"*

is the famous incipit of Schumann's review of Chopin's Op.2 Variations on "Là ci darem la mano" by Mozart) and he even included Chopin as one of the characters in his Carnaval (together with Paganini).

Finally, we would like to note that although Chopin always used only generic titles for his compositions (Études, Preludes, Nocturnes, Ballades, etc.) refusing any programmatic intent for his music, nevertheless many of his compositions have "attracted" titles, being very evocative of images and emotions. A list of some of these titles for études and preludes is reported in Table 1.

| Étude Op. 10 | Preludes Op.28 |
|---|---|
| n.1- Waterfall | n.1- Reunion |
| n.2- Chromatique | n.2- Presentiment of Death |
| n.3 - Tristesse | n.3- Thou Art So Like a Flower |
| n.4 - Torrent | n.4- Suffocation |
| n.5 - Black Keys | n.5- Uncertainty |
| n.6 - Lament | n.6- Tolling Bells |
| n.7 - Toccata | n.7- The Polish Dancer |
| n.8 - Sunshine | n.8- Desperation |
| n.11 - Arpeggio | n.9- Vision |
| n.12 - Revolutionary | n.10- The Night Moth |
| | n.11- The Dragonfly |
| Études Op. 25 | n.12- The Duel |
| n.1 - Aeolian Harp | n.13- Loss |
| n.2 - The Bees | n.14- Fear |
| n.3 - The Horseman | n.15- Raindrop |
| n.4 - Paganini | n.16- Hades |
| n.5 - Wrong Note | n.17- A Scene on the Place do Notre-Dame de Paris |
| n.6 - Thirds | n.18- Suicide |
| n.7 - Cello | n.19- Heartfelt Happiness |
| n.8 - Sixths | n.20- Funeral March |
| n.9 - Butterfly | n.21- Sunday |
| n.10 - Octave | n.22- Impatience |
| n.11 - Winter Wind | n.23 - A Pleasure Boat |
| n.12 – Ocean | n.24- The Storm |

Table 1: Some titles given to Chopin's Etudes (by A.Cortot) and Preludes (by H.von Bulow).

Apart from some obvious ones, referring to the technical aspect of a particular piece (thirds, sixths, octaves), the above names appear sometime to be very specific and in general completely arbitrary, especially in our times, in which there is a growing attention for the original (Urtext) version of compositions.

# 3 – Étude Op.10 n.3 – tonal symmetry breaking (with defects)

The Étude Op.10 n.3, also known as "Tristesse" (see above) is certainly one of the most famous among Chopin's compositions, for its lyrical character. The author himself is reported [6] to have said about it: "*In all my life I have never again been able to find such a beautiful melody*".

Such a theme appears in the first and third section of the étude, which has the structure of a song A-B-A', and indeed it is so beautiful to "obscure" the central section, which has a virtuoso character being based on progressions of double notes culminating with a sort of cadenza in double sixths in both hands.

In this respect it is interesting to read the review by C.Chaulieu, in Le Pianiste (n.1, 1833), about Op.10 n.3: *"The third one is among the most difficult ones. The melody is beautiful, but difficult to render. I recommend the chromatic scale in diminished sevenths which is new and of good taste; I cannot say the same for the following passage [the one marked con bravura] which has not a pleasing effect...."*

One thus is led to ask the reason of such a contrasting structure which may appear rather surprising and even disconcerting at a first sight. In the following, using very general mathematical (symmetry) arguments, we propose a possible key to understanding for this étude which could perhaps be useful for its interpretation.

Let us start with some very general considerations about our (Western) tonal system [8]: this is based on twelve notes equally spaced (equal temperament), namely a symmetric discretization of the natural interval (octave) corresponding to doubling of a given frequency, whereas tones with doubled frequency are perceived as the same. Consequently, tonal music deals with operations with 12 objects modulo 12 (cyclic group) [9,10]. From the point of view of perception, the affirmation of a given tonality correspond thus to a process of "symmetry breaking" among the otherwise *a priori* equivalent 12 tonalities (we do not distinguish here among major and minor modes). The idea of "symmetry breaking" is indeed one of the most fruitful in modern Physics, being associated to the emergence of ordered structures from the disorder (symmetry) [11].

In the specific case of Op.10 n.3, the tonality of E major appears in a very clear form in the main theme: we can think to this as to an ordered state (crystal) growing in time in front of us (see Fig.1).

Fig.1 Etude op.10 n.3 - Initial theme

Using Scriabin's correspondence between sounds and colors (see Appendix), this crystal would be blue/green. The crystal appears beautifully clear and perfect (Fig.2).

Fig.2 Etude op.10 n.3 first section – Ordered phase (blue crystal)

The second section, marked ``*più mosso*'', has a very different character: here soon appear some chromatic scales and dissonant sequences of double notes followed by a long passage in double-sixths, marked *con bravura* to remark its virtuoso character, and based on diminished seventh chords. These chords are symmetric ones, each made of four notes, equally spaced (see Fig.3). In our tonal system there are three such chords, exhausting the twelve notes.

Fig.3 Diminished seventh chords with their inversions. Note that the three cords cover all 12 notes.

The passage at bars 55-62 (see Fig.4) is a progression based on these three chords, developed over fixed intervals (sixths) and covering all 12 notes: it therefore represents a symmetric (disordered) phase - effectively atonal - although formally based on the dominant B+. Here the initial ordered state represented by the tonality E+ is destroyed: the blue crystal melts in the fire and all colors are present (symmetry restoration).

Fig.4. Etude op.10 n.3 – The symmetric phase (all colors are present). Symmetry breaking by repetition of C♮ (red notes in bar 62).

The return to the initial theme based on E+ tonality requires a breakdown of the tonal symmetry. This is realized at bars 61-62 (see Fig.4) by lowering the C# to C♮, thus altering the major sixth interval E-C# to E-C♮, to end finally on the dominant key B. At this point, a brief section of few bars (63-71) follows: here a "dissipation" of the large energy previously accumulated occurs, by means of irregular groups (triplets) in the bass, which give an impression of friction (slowing down) (Fig.5).

Fig.5. Etude op.10 n.3 – Transition episode (dissipation)

Finally, the original tonality of E major is reached and the theme reappears: the system has cooled down and the crystal has re-formed in the melting pot. However, this section is not an exact copy of the first one: some bars are missing, the dynamical marks are different (f instead of ff at bar 101), and especially, it appears a C♮ in the main theme at bars 104-105. This is very remarkable since such a note is not in the harmony of E+ key and produce the effect of adding a nostalgic feeling to the melody (justifying perhaps the title Tristesse). Note that this note is the same one which triggered the symmetry breaking at bar 62.

Coming back to the image of the crystal which has undergone melting and re-cristallization [12], we see that this is not anymore the same as the original, perfect, structure. Now some defects - red spots - appear in the blue gem (Fig.6). This vision is freely inspired by the Kibble-Zurek mechanism of topological defect formation in the course of symmetry breaking phase transitions [13]: such defects appear during the (fast) ordering process of the system and represent a "memory" of the symmetric phase. In the specific case, we have restoration and breakdown of tonal phase transition with the appearance of (harmonic) defects.

Fig.6. Op.10 n.3 third section – Ordered phase with harmonic defects (blue crystal with red spots)

At this point, admittedly, the reader could well complain: we have remarked (see Section 2) the arbitrariness of titles attached to Chopin's compositions and now we end up with a new image/title for this étude!

Actually, what we are proposing here is to associate to music a dynamical process (physical methapor), rather than a specific image. Indeed, in the same way as the Kibble-Zurek mechanism describes defect formation in a large number of physical systems (from cosmology to condensed matter), here the key point is the dynamics associated with the evolution of symmetry patterns, whereas the crystal melting and reforming with defects is a useful (by no means unique) visual tool.

We would like to stress that the real dynamical process should be thought to occur at level of perception (i.e. in our brains): the Artist, with his/her high sensitivity, is capable of giving a musical form to emotions, in a way that can be communicated to other people, by "resonating" in their brains as well.

By resorting to the above physical process, we finally propose a possible interpretation for the Étude Op.10 n.3: a metaphor of an internal trip from an adolescent age of unlimited, perfect, dreams - through some dramatic event - to a later stage, in which same emotions and dreams manifest in a necessarily different way. Within this framework, the central section of the étude, which appeared so disturbing to Chopin's contemporaries, plays a crucial rôle, hidden under a technical motivation.

As a final step in our reasoning, we can ask why the mathematical structure associated to physical processes should have something to do with our emotions (viewed as a part of brain dynamics). A possible answer is that physical processes describe well Music because musical structures are able to "excite" ordered structures already present in our brain (aesthetic experience), which are previously formed as images of natural phenomena. This would also explain why some music can be appreciated from a large number of people, independently of their cultural background.

In conclusions, the above considerations seem to suggest that Music can be used as an efficient tool, a "probe", for investigating brain dynamics, especially for what concerns emotions.

## 4 – Etude Op.25 n.1 – coherent states

This étude is known as "Aeolian harp", after the report about it by Robert Schumann, who listened it played by Chopin:

*"Imagine that an aeolian harp possessed all the musical scales, and that the hand of an artist were to cause them all to intermingle in all sorts of fantastic embellishments, yet in such a way as to leave everywhere audible a deep fundamental tone and a soft continuously-singing upper voice, and you will get the right idea of his playing.*

*But it would be an error to think that Chopin permitted every one of the small notes to be distinctly heard. It was rather an undulation of the A flat major chord, here and there thrown aloft anew by the pedal.*

*Throughout all the harmonies one always heard in great tones a wondrous melody, while once only, in the middle of the piece, besides that chief song, a tenor voice became prominent in the midst of chords. After the Etude a feeling came over one as of having seen in a dream a beatific picture which when half awake one would gladly recall."*

The technical aspect considered in this étude is indeed "arpeggios" which however are treated in a different way from other études on the same subject (eg Op.10 n.1 and n.11, Op.25 n.12). Here the arpeggios are developed in a compact way around a key note (upper or lower) and internal notes are written in small characters denoting, as testified by Schumann, the intention to give to such (small) notes only the function of "color" – an anticipation of impressionistic textures as for example in Debussy. This is visible in Fig.7 where the first few bars of the piece are shown.

Fig.7.  Etude op.25 n.1 – bars 1-4. Coherent states - unperturbed dynamics regime (see text).

The peculiar sound effect of this writing is that of "droplets" with an undefined internal structure. Thus it is tempting for a physicist to regard such "objects" as coherent states, with a well defined (macroscopic) envelop and an undefined (microscopic) particle content, as dictated by the phase-number uncertainty relation:

$$\Delta n \, \Delta \varphi \geq \tfrac{1}{2} \quad\quad\quad (1)$$

It is however to be remarked again that such a suggestive image should be rather thought to apply at level of perception than to the score itself. The meaning of Eq.1 is therefore the following: the distinct perception of the internal structure (small notes) is not compatible with the perception of the whole.

The étude proceeds in a rather uniform way, in an ecstatic atmosphere, according to Schumann's account: we have thus coherent states evolving with free, unperturbed, dynamics in the first part of the pièce.

Subsequently, however, a series of modulations produces a perturbed dynamics due to gradients in the harmony (potential) and the coherent states deform (squeezing): this is clearly visible in Fig.8 in the altered rhythmic structure of the left hand arpeggios (lower line).

Fig.8 Etude op.25 n.1 – bars 40-42. Coherent states - perturbed dynamics (squeezing) (see text).

The idea of the bass line acting as a sort of potential is supported by the analysis of the passage at bars 26-29 (Fig.9): here the upper voice moves twice up and down starting from E, and only the third time is able to reach the F# from which then proceeds further. This movement is accompanied by a progressive change in the harmony, which only at bar 28 provides sufficient energy to move away from the attractor E. This situation has a striking similarity with the behavior of a particle in a potential well, which inevitably goes back into the well unless sufficient energy is provided in order to pass the barrier: in some way, we could regard this passage as a "harmonic trap".

Fig.9. "Harmonic trap'': higher voice cannot reach F#, until sufficient energy is provided (by gradients in the harmony).

Towards the end, the dynamics relaxes again and, like in a dream, the whole image tends to vanish. It is very interesting what happens in the final bars 94-99 (see Fig.10): the lower A♭ "opens up" the closed chords in an arpeggio which extends up to the whole keyboard and dissipates the residual energy in a final trill. This process can be associated to decoherence and evaporation of the coherent states.

Fig.10. Etude op.25 n.1, final bars. Coherent states – Decoherence and evaporation.

In closing this Section, we would like to mention another example of "coherent states", namely Prelude op.28 n.8 (see Fig.11). Here, however, the arpeggios are not closed, contributing to the dramatic character of the piece (in a minor key, marked Agitato): these coherent states are the non-compact counterpart of those discussed above for the Etude op.25 n.1.

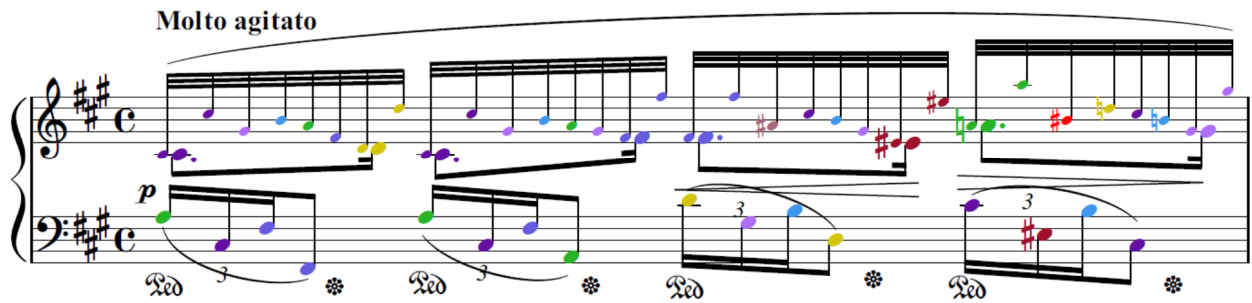

Fig.11. Prelude op.28 n.8 – Coherent states – non compact case.

# 5 – Conclusions and perspectives

In this paper we have explored the possibility of associating dynamical processes to Music with the aim of obtaining useful images (physical metaphors) for its interpretation. We did this in two specific cases, namely Chopin's Études Op.10 n.3 and Op.25 n.1: in the first case, we found a process of breakdown and restoration of tonal symmetry with the appearance of harmonic defects, in the spirit of the Kibble-Zurek mechanism. In the second case, the emerging dynamics is that of coherent states, undergoing time evolution and finally decoherence. We believe that this analysis could offer a new viewpoint on these compositions which hopefully can be of interest for musicians.

The dynamics we consider is generated by the harmonic relationships (forces) among different chords in the tonal system (an analysis of such relations is given in Refs.[14]). We have used colored musical scores, obtained by implementation of Scriabin's color code for sounds, to help visualizing such relations. The correspondence between musical dynamics and "forces" due to harmonic changes, viewed as gradients of a given potential, is one of the ideas we are currently investigating.

Another aspect we are considering is musical tempo, whose fluctuations could be related to the dynamical evolution of the harmonic patterns. One can thus have a "proper" time for the performer/listener as opposed to an "external" (metronomic) time. Other physical concepts which turn out to be relevant for our analysis are fractals, (harmonic) horizons and entanglement.

We have remarked how the physical processes we associate to Music should be really thought to take place at level of perception, i.e. in the brain. In this respect, the use made in our analysis of concepts from quantum physics well fits within the quantum description of some aspects of brain dynamics (consciousness) as proposed in Refs.[15-17].

## Acknowledgments

I thank the organizers of BEC2016 for the possibility of presenting these ideas in the form of a contribution to the Proceedings of the Conference. I especially thank Prof. R.Citro for the organization of the event (concert/seminar) and the encouragement to work on this subject. I also thank my wife Kateřina Fürstová for the collaboration as a musician and for performing together in various occasions, including the concert for this Conference.

Finally, I would like to remind that a first presentation of ideas contained in this paper (limited to Op.10 n.3), was given in December 2014 in London during a concert for the Conference DISCRETE-2014; I take this occasion to warmly thank the organizer of that event, Prof. N.E.Mavromatos.

# Appendix: Scriabin's color code for musical scores

In this Appendix we present an implementation of Scriabin's color code for tones to musical scores. This is done for being used in the main text in order to better emphasize the harmonic structures of the études which are discussed there. In alternative, the colored scores can be also useful for didactic purposes or simply for easier reading of music during study and/or performance.

The idea of associating colors to sounds is very old, dating back to ancient Greeks or even before. Newton established a correspondence among "fundamental" colors and the seven notes of the diatonic scale. Much later, a scheme based on the circle of fifths was proposed by Scriabin, who claimed to have synesthetic experience (visualizing colors while hearing sounds). This is reported in Fig.12: we see that similar colors are associated, for a given tonality, to the 4$^{th}$ and 5$^{th}$ degree of the scale, while distant tonalities have different color.

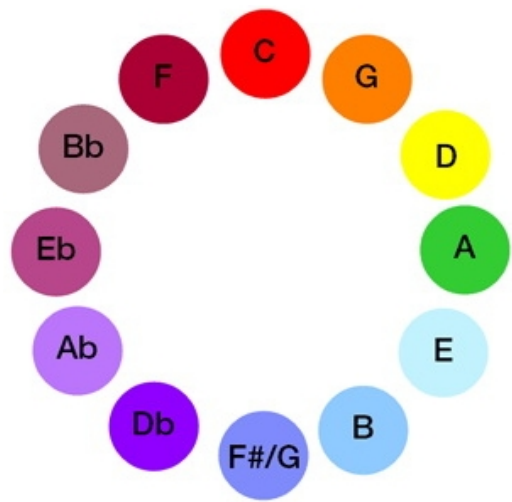

Fig.12. Colors arranged after Scriabin following the circle of fifths

Of course, the scheme only give relative values for the colors while the absolute value seems to be arbitrary (for example red for C). However, Scriabin claimed to "see" this and indeed he also elaborated a complex symbolism for the various colors/tonalities – for example:

C (Red) - Human will; D (Yellow) – Happiness; E (Light Blue/green) – Dream; F (Dark Red) – Creativity; A (Green) – Matter; B (Dark Blue) – Contemplation.

Once one accepts such a correspondence, then it is interesting to try to visualize this on the scores directly. It is possible to do this in a relatively easy way by means of the free software MuseScore (but also in other ways, for example with LaTeX – Lilypond). The result is as follows (Figs.13,14):

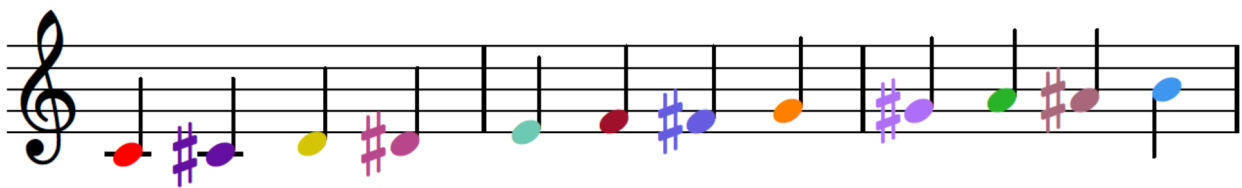

Fig.13. Realization of Scriabin's color code: chromatic scale.

Note that, due to the small size of the notes, some of the colors of Fig.12 needed to be slightly darkened in order to be properly appreciated on the white background.

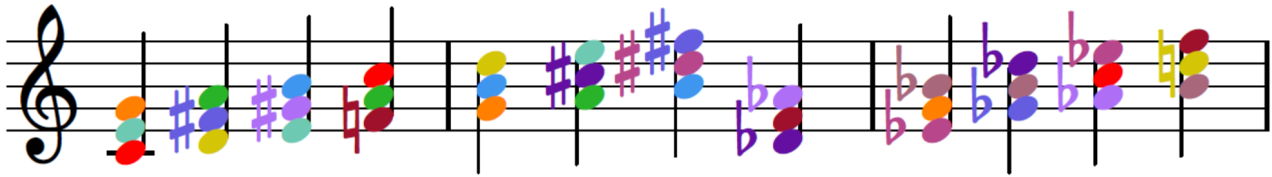

Fig.14. Realization of Scriabin's color code: major triads.

Finally, we observe that colors have been already used in literature for analysis of musical text, see for example Ref.[18].